\documentclass[aps,superscriptaddress,showpacs,nofootinbib]{revtex4}
\usepackage{epsfig}

\begin{document}

\title{Exact $1/N$ and Optimized Perturbative
Evaluation of $\mu_c$ \\for Homogeneous Interacting Bose Gases}

\author{Jean-Lo\"{\i}c Kneur}
\email{kneur@lpm.univ-montp2.fr}
\affiliation{Laboratoire Physique Math\'{e}matique et Th\'{e}orique - CNRS -
UMR
5825 Universit\'{e} Montpellier II, Montpellier, France}

\author{Marcus B. Pinto}
\email{marcus@fsc.ufsc.br}
\affiliation{Departamento de F\'{\i}sica,
Universidade Federal de Santa Catarina,
88040-900 Florian\'{o}polis, SC, Brazil}

\begin{abstract}

In the framework of the $O(N)$ three-dimensional
effective scalar field model for homogeneous 
dilute weakly interacting Bose gases
we use the $1/N$ expansion,
within the large $N$ limit, to evaluate
 the parameter $r_c$ which is  directly related to the
critical chemical potential $\mu_c$.
This quantity enters the order-$a^2 n^{2/3}$
coefficient contributing to the critical temperature shift
$\Delta T_c$ where $a$ represents the $s$-wave scattering length and 
$n$ represents the density. Compared to the recent precise 
numerical lattice simulation results,
our calculation 
suggests that the large $N$ approximation performs rather
 well even for the physical case $N=2$. We then 
calculate the same quantity but using different forms of    
the optimized perturbative 
(variational) method, showing that these produce
excellent results
both for the finite $N$ and large-$N$ cases.

\end{abstract}

\maketitle

\section{Introduction}

An issue that has attracted considerable attention recently and
that is associated to the perturbation theory breakdown problem is the
study of how interactions alter the critical temperature ($T_c$) of
Bose-Einstein condensation (BEC). Due to its nonperturbative nature,
this is clearly a non-trivial problem.
The authors of Ref. \cite{second} were
able to show that the transition temperature for a dilute, homogeneous,
three dimensional Bose gas can be expressed at next to leading order as

\begin{equation}
\frac { \Delta T_c} {T_0} =  c_1 a n^{1/3} + \left[ c_2^{\prime} \ln(a n^{1/3})
+c_2^{\prime \prime} \right] a^2 n^{2/3} + {\cal O} \left(a^3 n \right)\;,
\label {exptc}
\end{equation}
where $T_0$ is the ideal gas critical temperature, $a$ is the $s$-wave scattering
length and
$n$ the number density.
A similar structure is also discussed in Ref. \cite {markus}. As far as the
numerical coefficients are concerned, the exact value, $c_2^{\prime}=-64
\pi \zeta(1/2) \zeta(3/2)^{-5/3}/3 \simeq 19.7518$, was obtained using
perturbation theory \cite {second}. On the other hand, the other two
coefficients, $c_1$ and $c_2^{\prime \prime}$, are sensitive to the
infrared sector of the theory and consequently cannot be obtained
perturbatively, but they can, through a matching calculation, be
expressed in terms of the two nonperturbative quantities $\kappa$ and
${\cal R}$ which are, respectively, related to the number density
 and to the critical chemical potential, 
as discussed in the next section. 
Some early analytical predictions for $c_1$
included the self-consistent resummation schemes  ($c_1 \simeq 2.90$) \cite
{baymprl}, the $1/N$ expansion
at leading order ($c_1 \simeq 2.33$) \cite {baymN} and at next to leading order
($c_1 \simeq 1.71$) \cite {arnold1} 
and also
the linear $\delta$ expansion (LDE) at second order ($c_1 \simeq 3.06$) \cite
{prb}. Today the most precise
values predict $c_1 \simeq 1.30$
as shown by the lattice (Monte Carlo) simulations  which produce $c_1 = 1.29 \pm
0.05$ \cite {russos} and
$c_1 = 1.32 \pm 0.02$ \cite {arnold2,arnoldrc} whereas
 very recent analytical
studies, up to seven loop,
predict $c_1 = 1.27 \pm 0.05$
(Variational Perturbation Theory, \cite {kastening2}) and $c_1 = 1.30$ (improved
LDE, \cite {knp}). A
clear review of all how those methods apply to the
problem can be found in Ref. \cite {andersen}.
At the same time, the order-$a^2 n^{2/3}$ non-perturbative
 coefficient, $c_2^{\prime \prime}$, has not been considered by most
authors working on the BEC $\Delta T_c$ problem. At this point it is worth
recalling that
while $c_1$ is a function only of $\kappa$, which is related to the number
density, $c_2^{\prime \prime}$
is a function of $\kappa$ as well as ${\cal R}$, which is related to the critical
chemical potential. 
This coefficient was first evaluated with
lattice simulations which predict $c_2^{\prime \prime}= 75.7 \pm 0.4$ for $N=2$
\cite {second} and
a previous analytical  evaluation made use of the standard
LDE \cite{pra,new} as well as its recently improved version \cite
{knp}.
The best previous analytical results (in comparison
to the lattice ones) were produced with the improved LDE which obtains 
$c_2^{\prime \prime}\simeq 73.5$  at order-$\delta^5$ \cite {knp}. 
Very recently, the highest presently available (seven-loop order) 
perturbative contributions and a 
corresponding new evaluation of  ${\cal R}$ within the framework of 
the above mentioned variational perturbation theory, has been performed
by Kastening \cite{newkast}.
The purpose of the
present work is twofold.
First, we use the traditional $1/N$ expansion to evaluate analytically, 
in the large-$N$ limit,
the above mentioned quantity $ {\cal R}$. 
This large $N$ result is a useful cross-check of a similar evaluation
also presented very recently in Ref. \cite{newkast}, since our
independent calculation is performed in  
a quite different way, as
discussed in detail below. 
Using
this result in conjunction 
with the results of Ref. \cite {baymN} for 
the density number one may finally
establish the analytical large-$N$ prediction for
$c_2^{\prime \prime}$
at the next-to-leading $1/N$ order. 
Comparing our large-$N$ results for $\mu_c$ with
the ones
produced by the lattice simulations we show how 
the $1/N$ approximation performs very well, already
at the first non trivial order, even
for the homogeneous Bose gas case where $N=2$.
Second, by considering our $1/N$ result as being ``exact"
we perform standard as well as improved (LDE) optimized perturbative 
evaluations of ${\cal R}$ (or equivalently $\mu_c$),
 also at the $1/N$ order, so that the convergence and reliability
of such a method in the BEC context can be further tested.
As we recall below, the LDE method essentially introduces a  
mass term within the effective scalar field model, therefore avoiding the 
above mentioned infrared divergence problems of ordinary 
perturbation theory. Then, the interest of considering 
LDE in the large-$N$ limit is that
at the relevant (next-to-leading) $1/N$ order we can define
and calculate the LDE perturbative series to {\em arbitrary} 
orders for the quantity $r_c$, and thus check the convergence
of the LDE method in principle to arbitrary orders.
 Our optimized
perturbation analysis, 
in the large-$N$
limit, shows how the improved
versions of this method also perform very well 
in connection with the $\Delta T_c$
problem. \\
This work is organized
as follows. In the next section we briefly recall the
effective model which is considered throughout the paper. 
In  Section III we
perform the  exact analytical evaluation at $1/N$ order of
the critical chemical potential.
Section IV is devoted to the optimized
perturbation (LDE) applications both for the large-$N$
as well as the finite $N$ case.
We pay  special attention to the convergence problem. Our
conclusions are presented
in Section V. 

\section { The effective scalar three dimensional model}

The studies concerning the equilibrium properties of BEC can be
addressed by means of a non-relativistic effective theory described by a
complex scalar field. In the dilute limit, which is the regime involved
in those experiments, only two-body interactions are important
\cite{second} and one may then consider the following $U(1)$ invariant
finite temperature Euclidean action

\begin{equation} S_E = \int_0^\beta d\tau \int d^3 x \left\{
\psi^*({\bf x},\tau)\left( \frac{d}{d\tau}-\frac{1}{2m_A}\nabla^2\right)
\psi ({\bf x},\tau) -\mu \psi^* ({\bf x},\tau) \psi ({\bf x},\tau) +
\frac{2 \pi a}{m_A} \left[\psi^*({\bf x},\tau)\psi({\bf x},\tau) \right]^2
\right\}\;,
\label{SE}
\end{equation}
where, in natural units, $\beta$ is the inverse of
the temperature, $\mu$ is the chemical potential
and $m_A$ the mass of the atoms. At the relevant low
temperatures involved in BEC the internal degrees of freedom are
unimportant and this can be taken as an effective model of hard core
spheres with local interactions for which $a$ represents the $s$-wave
scattering length.

The field $\psi$ can be decomposed into imaginary-time frequency modes
$\psi_j ({\bf x},\omega_j)$, with discrete bosonic Matsubara frequencies
$\omega_j = 2 \pi j/ \beta$, where $j$ is an integer. Near the
transition the chemical potential becomes very small as compared to the
temperature ($|\mu|<<T$) and, since the correlation length tends to
infinity, the distances are large compared to the thermal wavelength
$\lambda= \sqrt{2\pi\beta/m_A}$.
Therefore, only the zero modes 
matter and one may integrate out all other modes to
obtain
an effective three-dimensional scalar action 
with an $O(2)$ symmetry, given by \cite {baymprl}
\begin{equation}
S_{\phi}=  \int d^3x \left [ \frac {1}{2} | \nabla \phi |^2 +
\frac {1}{2} r
\phi^2 + \frac {u}{4!} (\phi^2)^2 + \frac {A}{2} \phi^2
\right ] \;,
\label{action2}
\end{equation}

\noindent
where  $r =- 2m_A \mu$ and $u=48 \pi a m_A T $.  
In the large-$N$ limit mainly considered in
this work, and also in Refs. \cite {baymN,arnold1,braaten,new}, the
field $\phi$ in Eq. (\ref{action2}) is formally considered as having $N$
components ($\phi_i$, $i=1,\ldots,N$). In this case, the Bose-Einstein
condensate effective action Eq. (\ref{action2}) is the $N=2$ special
case of the general $O(N)$ invariant action. The 
three-dimensional
effective theory described by Eq. (\ref {action2})
is super-renormalizable and requires only a mass counterterm ($A$) to
eliminate any ultraviolet divergence.
Using this action one may use any non perturbative technique to evaluate the
physical
quantities related to the number
density
\begin{equation}
\kappa \equiv \frac {\Delta \langle \phi^2 \rangle_c}{u}  =
\frac { \langle \phi^2 \rangle_u - \langle \phi^2 \rangle_0}{u} \;,
\label{kappa}
\end{equation}
and to the critical chemical potential

\begin{equation}
{\cal R} \equiv \frac {r_c}{u^2} = -\frac {\Sigma(0)}{u^2}  \;,
\label{R}
\end{equation}
where the subscripts $u$ and $0$ in Eq. (\ref{kappa}) mean that the
density is to be evaluated in the presence and in the absence of
interactions, respectively, and $\Sigma(0)$ is the self-energy with zero
external momentum.

 The actual relations between the two
nonperturbative coefficients appearing in Eq. (\ref {exptc}) and these physical
quantities are given by
\cite {second}

\begin{equation}
c_1 = - 128 \pi^3  [\zeta(3/2)]^{-4/3}  \kappa   \;\;,
\label{c1}
\end{equation}
and

\begin{equation}
c_2^{\prime \prime} = - \frac{2}{3} [\zeta (3/2)]^{-5/3} b_2^{\prime \prime} +
\frac {7}{9} [\zeta (3/2)]^{-8/3} (192 \pi^3 \kappa)^2 + \frac{64 \pi}{9}
\zeta (1/2) [\zeta(3/2)]^{-5/3}
\ln \zeta (3/2)   \;,
\label{c2}
\end{equation}
where $b_2^{\prime \prime}$, in Eq. (\ref{c2}), is

\begin{equation}
b_2^{\prime \prime} = 32 \pi \left \{ \left [ \frac{1}{2} \ln (128 \pi^3) +
\frac{1}{2} - 72 \pi^2 {\cal R} - 96 \pi^2 \kappa \right ]\zeta(1/2)
 + \frac {\sqrt {\pi}}{2} - K_2 - \frac {\ln 2}{2 \sqrt {\pi}}\left [
\zeta(1/2) \right ]^2 \right \}  \; ,
\label{b2primeprime}
\end{equation}
with $K_2= -0.13508335373$.

As discussed below, the
relation between $r_c$ and $\Sigma(0)$ comes from the Hugenholtz-Pines
(HP) theorem at the critical point.
As a matter of fact, when quantum corrections are taken into account, the full
propagator for the
effective three-dimensional theory reads

\begin{equation}
G(p)=[p^2+ r +\Sigma_{\rm ren}(p)]^{-1} \;,
\end{equation}
where $p^2$ represents the three-momentum and $\Sigma_{\rm ren}(p)$
represents the renormalized self-energy. At the transition point
($p^2=0$), the system must have infinite correlation length and one then
has

\begin{equation}
[G(0)]^{-1}=[r_c +\Sigma_{\rm ren}(0)] =0 \;.
\end{equation}
This requirement leads to the Hugenholtz-Pines theorem result
$r_c=-\Sigma_{\rm ren}(0)$. Since $r_c$ is at least of order $u$, it
would be treated as a vertex in a standard perturbation type of
calculation in which $G(p)=1/p^2$ represents the bare propagator. This
shows that perturbation theory is clearly inadequate to treat the BEC
problem at the transition due to the presence of infrared divergences.
One must then reccur to nonperturbative methods like the numerical
lattice Monte Carlo simulations (LS), the analytical $1/N$ technique or the LDE
\cite {linear}  (see for
instance Refs. \cite{early} for different types of applications).
The problem is highly nontrivial since the Hugenholtz-Pines theorem
automatically washes out all momentum independent contributions, such as
the one-loop tadpole diagrams, which constitute the leading order of
most approximations. In practice, this means that the first nontrivial
contributions start with two-loop momentum dependent self-energy terms.
However, having reduced the original model, Eq. (\ref{SE}), to the
effective three-dimensional one, Eq. (\ref{action2}),
makes it easier to tackle those
contributions since one no longer has the problem of summing over the
Matsubara's frequencies, which is a hard task when the number of loops
increases.

Here, as stated in the introduction, our aim is to apply the $1/N$ expansion and
the LDE to Eq. 
(\ref {action2}) to explicitly evaluate the quantity $r_c= - 2m_A \mu_c$ 
($m_A$ being the relevant atom mass), comparing
our results with the
the lattice simulation prediction which, for $N=2$, is $r_c^{\rm LS} = (0.0019201
\pm
0.0000021) u^2$ \cite {second, arnoldrc}.

\section{Standard Large-$N$ Calculation}

The large-$N$ evaluation is carried out, as usual, by assigning a 
factor of $N$ to any closed loop, and 
considering $u \sim 1/N$.
Then for the
massless critical theory given by Eq. (\ref {action2})
the first non trivial contributions 
are at the next-to-leading order $1/N$, and
come from the tadpole (with bubble
insertions) 
type of graphs (see Figure 1):
\begin{figure}[htb]
\epsfig{figure=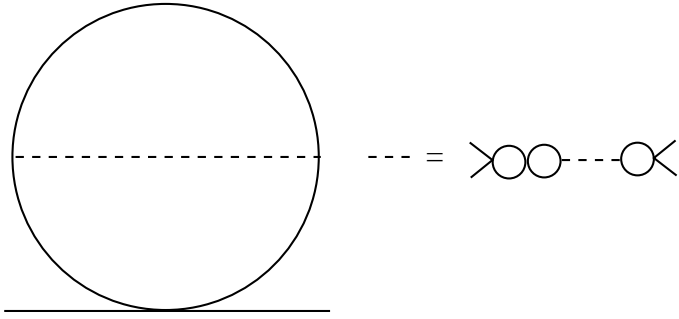,angle=0,width=9cm}
\caption[]{\label{fig1a} Feynman graphs for the tadpole
 contributions
to $r_c$ at $1/N$ order, with the resummed propagator
(dotted lines).}
\end{figure}
\begin{figure}[htb]
\epsfig{figure=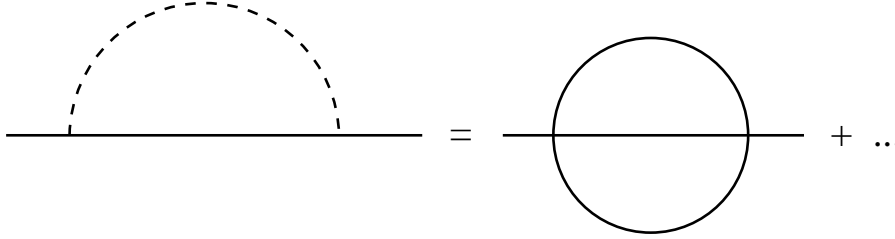,angle=0,width=12cm}
\caption[]{\label{fig1b} Feynman graph for the chain contribution
to $r_c$ at $1/N$ order.}
\end{figure}
\begin{equation}
r_c^{(N,{\rm tad})}
= - \frac {N}{6} u  \int \frac{d^3\,p}{(2\pi)^3} \;
 \frac{1}{p^2 +\Sigma(p)-\Sigma(0)} \;,
\label{basm0}
\end{equation}
where
\begin{equation}
\Sigma(p) =\frac{2}{N} \int \frac{d^3\,k}{(2\pi)^3}\;
\;F(k)\;\frac{1}{(k+p)^2}  \;\;,
\end{equation}
with the ``dressed" (resummed) scalar propagator

\begin{equation}
F(k) = \left (\frac{6}{N\,u} +B(k) \right )^{-1}   \;\;,
\end{equation}
where

\begin{equation}
B(k,D) = \int \frac{d^D\,q}{(2\pi)^D} \;\frac{1}{q^2\;({ k} +{ q})^2}   \;
\to \frac{1}{8\:k}\;\; \mbox{for}\; D \to 3,
\label{BkD}
\end{equation}
represents the basic one-loop $D$-dimensional (massless) integral. 
The evaluation of Eq. (\ref{basm0}) 
can be done exactly and most conveniently 
in dimensional regularization. Actually,
 up to a trivial factor of $-u/6$, it is exactly the same
two-loop integral entering the exact evaluation
of $\langle \phi^2 \rangle$ at the $1/N$ order,
thus already performed in Ref. \cite{baymN}
(see also Appendix A of Ref.~\cite{new} for details). The result
(which is finite in three dimensions) is

\begin{equation}
r_c^{(N,{\rm tad})} = \frac {N}{6} \frac {u^2}{96 \pi^2} \,\,.
\label{rctad}
\end{equation}

The other contributions to $r_c$
at the next-to-leading $1/N$ order come from the chain exchange 
type of graphs (see Figure 2):

\begin{equation}
r_c^{(N,{\rm exc})} = -\Sigma^{(N,{\rm exc})}(0) =
- \frac{2}{N} \int \frac{d^3\,k}{(2\pi)^3}\;
\;F(k)\;\frac{1}{k^2}  \;\;,
\label{sigma}
\end{equation}
which can also be calculated exactly in dimensional regularization
in a very similar way. The result, however, is not finite in three
dimensions 
 and has therefore to be regularized and renormalized. 
Although it turns out to be a technically
rather standard two-loop calculation in dimensional regularization,
we shall give it in some details below since, to the best of
our knowledge, it has not been done before. 

Due to the non-finiteness of the final result
one should use the arbitrary $D$-dimensional result 
for the propagator in Eq.
(\ref{BkD}) which reads
 \begin{equation}
B(k,D) = \frac{2^{3-D} \sqrt{\pi}}{(4\pi)^{D/2}}\;
\frac{\Gamma(D/2-1)\,\Gamma(2-D/2)}{\Gamma(D/2-1/2)}\;(k^2)^{D/2-2}
\equiv b(D)\:(k^2)^{D/2-2}\;.
\label{BkDres}
\end{equation}
Taking this result in the ($D$-dimensional) integral, Eq. (\ref{sigma}),
and making a convenient change of variable, $k = P^{2/(D-4)}$, one obtains
 the following integral:

\begin{equation}
\Sigma^{(N,{\rm exc})}(0) = \frac{4}{N\,b(D)}\:\frac{1}{4-D}\:\int
\frac{d^D P}{(2\pi)^D}\;
\frac{(P^2)^{(\frac{2-D}{4-D}-D/2)} }{P^2 +\frac{6}{N\,u \,b(D)}} \,\,,
\end{equation}
which we wrote in a form convenient for straightforward dimensional
regularization integration. The explicit result is 
\begin{equation}
\Sigma^{(N,{\rm exc})}(0) =
\frac{4\:\Gamma(\frac{2-D}{4-D})\Gamma(1 +\frac{D-2}{4-D})}
{(4\pi)^{(D/2)}\,(4-D)\:N\,b[D]\:\Gamma(D/2)}\:
 \left (\frac{6}{b(D)\,N\,u} \right )^{(-1-\frac{D-2}{4-D})}\;.
\label{sigD}
\end{equation}
Taking now the limit $D\to 3$ 
and taking into account the standard $\overline{\rm MS}$
scheme conversion factor:
  \begin{equation}
u \to u  \;\left (\frac{e^{\gamma_E/2}\: M}{2 \sqrt{\pi}} \right )^{3-D}\;,
\end{equation}
where $M$ is the arbitrary renormalization scale and 
$\gamma_E \simeq 0.51776$ the Euler constant one obtains,
 after renormalization by the appropriate counterterm (A) which removes
the $-N\,u^2/(6 \times 96\pi^2 (3-D))$ divergence, the final result:

 \begin{equation}
r^{(N,{\rm exc})}_c \equiv -\Sigma^{(N,{\rm exc})}(0) = \frac{N}{6} \frac
{u^2}{96\pi^2}
\:\left [1 +2\:\ln \left (\frac{48\, M}{N\,u}\right )\:\right ]\;.
\label{rcexc}
\end{equation}
 Numerically, for the standard lattice simulation choice
of scale, $M =u/3$, and the 
physically relevant number of field components, $N=2$, 
this gives
$r^{(N,{\rm exc})}_c \simeq 0.001814945\; u^2$.\\
Then, adding the two contributions, Eqs. (\ref{rctad})
and (\ref{rcexc}), one gets
 \begin{equation}
r_c^{(N)} = \frac{N}{6} \frac {u^2}{48\pi^2}
\:\left [1+\ln\left (\frac{48\,M}{N\,u}\right )\:\right ]\;,
\label{rcNex}
\end{equation}
which is completely equivalent to the result very recently 
obtained in Ref. \cite{newkast} (Eq. (25)), the latter being performed in
a different independent way.  
Eq. (\ref{rcNex}) is directly comparable with 
the available lattice simulation
results, since the latter were also obtained \cite{second, arnoldrc}
in the $\overline{\rm MS}$ scheme.\\ 
As an interesting consistency check, note that the logarithmic
scale dependence in Eqs. (\ref{rcNex}), (\ref{rcexc}) is
just identical to the one already obtained \cite{second}
 as the coefficient $c^\prime_2$ in Eq. (\ref{exptc}): more precisely,
putting the appropriate factors defined in Eq. (\ref{c2})
to pass from $r_c$ to the relevant coefficients of the expression
(\ref{exptc}) (and also putting back the arbitrary $N$
dependence of the relevant graph instead of its large $N$ 
value in Eq. (\ref{rcexc})), we have:
$c_2^{\prime\prime}(\ln {\rm part}) = 16(N+2) \pi\:
\zeta(1/2) \zeta(3/2)^{-5/3}/3\;\ln[48M/(N\,u)]$.  
This is of course expected, 
since the divergence has a perturbative origin and arises
solely from the setting sun graph, namely the order $u^2$ in
a formal perturbative expansion of expression (\ref{sigma}) (see
also Fig. 2).  In that sense, the   
novelty of our result resides more in the non-logarithmic
piece of Eq. (\ref{rcNex}), as well as the explicitly non-perturbative
dependence on the coupling $u$ therein.  

For $N=2$ and $M=u/3$ we obtain numerically the exact large-$N$ result
$r_c^{(N)}\simeq 0.002166755\, u^2$ which, when used in conjunction with the value
$\kappa=-Nu/(96\pi^2)$ obtained
by Baym, Blaizot and Zinn-Justin \cite {baymN} gives from
Eq. (\ref{c2}):
\begin{equation}
c_2^{\prime \prime}(\mbox{$1/N$ order, $N=2$}) \simeq 
95.7941\;.
\end{equation}

\section {Optimized perturbation evaluation}

In this Section our aim is to test the eventual convergence of the 
optimized perturbation (LDE method)
towards
the ``exact" large-$N$ result for $r_c$ given by Eq. (\ref{rcNex}).
This supplements a similar analysis performed 
in Ref. \cite{knp} for the quantity $c_1$ in the large-$N$ limit,
but for the independent quantity $c^{\prime\prime}_2$, so as to 
establish the reliability
of the finite $N$ LDE results recently obtained in Ref. \cite {knp}.
According to the usual prescription the LDE can be
implemented in the BEC case by considering the interpolated action
\begin{equation}
S_{\phi}^{\delta}=  \int d^3x \left [ \frac {1}{2} | \nabla \phi |^2+
\frac{1}{2}m^2 \phi^2 +
\frac {\delta}{2} (r -m^2)
\phi^2 + \delta \frac {u}{4!} (\phi^2)^2 + \delta \frac {A_{\delta}}{2} \phi^2
\right ] \;,
\label{actiondel}
\end{equation}
which interpolates between a free
theory ($\delta=0$) and
the original interacting theory ($\delta =1$) \cite {prb,pra,new}. Any physical
quantity ($\Phi$) is then evaluated and renormalized exactly
at some perturbative order  by treating the dummy parameter ($\delta$) as a small
number and $m$ as an arbitrary mass parameter to be fixed later. At the end,
one sets $\delta=1$. 
While the $\delta m^2$  is treated as a
new quadratic vertex,
the other part is absorbed by the propagator which is now
naturally infrared
regulated by the mass term. Accordingly, the LDE amounts in a first
stage to perform an order by order perturbative calculation of the relevant
Green's functions (here the self-energy contributions $\Sigma(0)$)
but in the massive theory as defined by Eq. (\ref{actiondel}).  
Thus, any finite
order perturbative calculation will produce $m$-dependent 
physical quantities. To
get rid of this unwanted dependence one usually 
demands that $\Phi$ has its optimum value at the point where it is less
sensitive to $m$:
\begin{equation}
\frac {d \Phi}{dm}=0 \;,
\end{equation}
according to the Principle of Minimal Sensitivity (PMS) \cite
{pms} to which we shall refer as ordinary PMS (OPMS).
However, we shall also consider below some specific variations 
of this method  that had been more recently 
developed\cite{kleinert,knp}, which in practice 
essentially introduce some additional arbitrary variational parameters. 
As explicitly shown in Refs. \cite {marcus1,gn2},
the whole procedure
is compatible with the
renormalization program which is implemented exactly as in the usual perturbative
case.
Finally note that,
 according to the original prescription \cite {prb}, we have treated $r$ ($r_c$ at
the critical point) 
 as an interaction, since this quantity has a critical value which 
is at least of
order $\delta$. The Feynman rules for the
 interpolated theory, in Euclidean space, are $-\delta r$, $\delta m^2$ and
$-\delta A_{\delta}$, for the quadratic vertices and 
 $- \delta u$ for the quartic vertex while the propagator is given by
 \begin{equation}
 G(p) = [p^2 +m^2]^{-1} \;\;.
 \end{equation}

\subsection {Reviewing the perturbative LDE calculation}

To make this paper self contained let us briefly recall how the LDE is employed in
the evaluation of $r_c$
by redoing the step by step evaluation  performed in Ref. \cite {pra}. Here, we
shall stay within the 
large-$N$ limit only, while the evaluation performed in Ref. \cite {pra} was
carried out for any value of
$N$.
According to the HP theorem, applied in conjunction with the LDE
\cite {prb,pra}
, $\delta r_c^{(\delta)}$ is obtained from the perturbative evaluation  of self
energy diagrams,
$\Sigma^{(\delta)}(0)$. To order-$\delta$ one has only the plain tadpole
contribution and
a direct application of the 
Feynman rules for the interpolated theory and dimensional regularization (see Ref.
\cite {pra} for details)
gives, in the large-$N$ limit, the finite contribution
\begin{equation}
\delta r_c^{(1)} = - \Sigma^{(1)}(0) = \delta u \frac {N}{3} \frac{m}{8\pi} \;\;.
\label{tad}
\end{equation}
Carrying on to order-$\delta^2$, still within the large-$N$ limit, one must 
consider \cite {pra}

\begin{equation}
\delta r_c^{(2)}= -\Sigma^{(2)}(0) = \delta u \frac {N}{3}
\frac{m^*}{8\pi} + \delta^2 \frac {N}{3} \frac {u}{16 \pi}
\frac {r_c}{m} 
- \delta^2 \left ( \frac{N}{3} \right )^2 \frac {u^2}{128 \pi} \;\;,
\label{delrc}
\end{equation}
where $m^*=m(1-\delta)^{1/2}$ must be expanded accordingly.
Now, one replaces $\delta r_c$ appearing in the right hand side
of the above
equation, 
with the value $\delta r_c^{(1)}$ obtained at the previous order so that the right
hand side {\it remains} of
order-$\delta^2$. Then, one can   see how the
``double-scoop" contribution
(third term on the RHS of Eq. (\ref {delrc})) is exactly canceled due to the HP
condition applied to $r_c$ at first order. It is easy to convince oneself that at
any higher order ($n$) 
all large-$N$
contributions of order-$(N\, \delta u )^k$, with $ 1 < k \le n$ cancel. Therefore
to get non trivial
results with the LDE one must add the next-to-leading order, $O(N^k \, (\delta u
)^{k+1})$ terms,
 shown in Figure 1, to Eq. (\ref {tad}).
The first comes from the ``setting-sun" contribution 
(see left-hand-side of equality in Figure
2),
which has been explicitly evaluated by Braaten and Nieto \cite {branieto}
and also in Ref. \cite {pra} giving

\begin{equation}
-\Sigma_{\rm ssun}(0)= \delta^2 \frac {N u^2}{18 (4\pi)^2}\left
[\frac{1}{4\epsilon} +\ln\left
(\frac{M}{m} \right ) 
+\frac{1}{2} +\ln(1/3) \right ] \;.
\end{equation}
As already mentioned at the end of section III,
this contribution displays
an ultraviolet pole as $\epsilon \equiv (3-D)/2 \to 0$ ($D\to 3$)
which is, within dimensional
regularization, the only primitive 
ultraviolet divergence associated with the effective 
super-renormalizable three-dimensional theory.  The pole
produced by this primitive
divergence fixes the mass counterterm 
coefficient in the $\overline {\rm MS}$
scheme
\begin{equation}
\delta A_{\delta} = - \delta^2 \frac {N}{18} \frac {u^2}{(8\pi)^2} \frac
{1}{\epsilon} \;\;.
\end{equation}
As usual, this ``vertex" must be considered also at higher orders so diagrams
whose divergences arise from 
``setting-sun" sub-diagrams maybe rendered finite.
One then gets the finite second order result
\begin{equation}
\delta r_c^{(2)}= -\Sigma^{(2)}(0) = \delta u \frac {N}{3} \frac{m^*}{8\pi}  +
\delta^2 \frac {N u^2}{18 (4\pi)^2}\left [\ln\left (\frac{M}{m} \right )
+\frac{1}{2} +\ln(1/3) \right ] \;\;.
\label{delrcrenor}
\end{equation}
Note that contrary to the original massless case,
 the massive LDE interpolated
action allows for the unique 
plain  tadpole
contribution, given by Eq. (\ref {tad}), to survive at any order. 
This $O(N^0)$
term must then appear
together with all $O(N^{-1})$ non trivial terms shown in Figures 1, 2.
In fact, as far as the LDE is concerned,  the linear term is crucial for 
convergence
properties \cite {new}.
Finally, the relation between the values of $r_c^{(\delta)}$
evaluated at two different 
($\overline {\rm MS}$) mass scales, $M_1$ and $M_2$, 
can be readily obtained from
Eq.
(\ref {delrcrenor}) and
reads
\begin{equation}
r_c(M_1)=r_c(M_2) + \frac {N u^2}{18(4\pi)^2}
\ln \left ( \frac {M_1}{M_2} \right ) \,\,.
\end{equation}
It is not too difficult to see that this relation will be verified at any order in
$\delta$. At order-$\delta^2$ the only
diagram which is scale dependent is the ``setting sun". At a higher order ($n \ge
3$) 
this order-$\delta^2$ type of contribution can appear only as a subdiagram
($\Sigma_{\rm ssun}(p)$). At the same
order a similar graph appears, but this time $\delta r_c$
replaces the setting sun insertion. However, the ``vertex" $\delta r_c$ is always
replaced by its expansion in $\delta$
which contains, at order-$\delta^2$, exactly the same scale dependent term as
given by the setting sun with zero external momentum ($\Sigma_{\rm ssun}(0)$),
with a
reversed sign producing a net combination proportional to $\Sigma_{\rm
ssun}(p)-\Sigma_{\rm ssun}(0)$
which, in three dimensions, turns out to be scale independent \cite {pra}. This
means that, apart from the order-$\delta^2$ setting sun, all
contributions to $\delta r_c^{(n)}$ 
are automatically scale independent. This perturbative procedure, to
order-$\delta^4$ is clearly illustrated
 in Ref. \cite {pra}. In the present work we are interested in the large-$N$ limit
only so it is easier to add up the whole series (represented by the diagrams of
figures 1 and 2)
 in a closed form and then develop it to the desired order in $\delta$ as
illustrated in the  next subsection.

\subsection{Resumming the large-$N$ series}

As shown in the previous subsection,
 a main difference between the LDE and
the exact large-$N$ above calculation is that with the former one
considers a massive theory, which effectively introduces a
mass term in the (dressed) propagator. So, for practical reasons the LDE
evaluation can be 
performed as discussed in the previous sub-section 
or alternatively as we do
now. 
Having defined the relevant perturbative series in powers of $u$ for the massive
case,
we can proceed to the usual ordinary PMS (OPMS)  or its improved version (IPMS).
Accordingly,
we apply
on the latter series the substitutions:
\begin{equation}
m \to m^*= m\,(1-\delta)^{1/2}\;;\;\;\; u\to u\, \delta \;\;,
\label{subst}
\end{equation}
in the standard OPMS-LDE \cite {linear}, or
\begin{equation}
m \to m\, (1 - \delta)^{1/2} [ 1 + (1-a)\, \delta ]^{1/2}\;;\;\;\;
 u\to u\, \delta
\;\;,
\label{subst2}
\end{equation}
and subsequent generalizations in the IPMS-LDE \cite{knp}. 
The difference between
the OPMS and the 
IPMS is basically that within the latter one introduces
more arbitrary interpolation parameters at each successive
perturbative order
such that  the PMS criterion is generalized by requiring
$\partial \Phi /\partial m =0$,
 $\partial^2 \Phi /\partial m^2 =0$, etc.
It turns out \cite {knp} that the 
IPMS essentially reduces a basic problem 
concerning the OPMS. In fact, as discussed in many BEC applications \cite {pra,
braaten, new}, at high orders 
the OPMS generates many possible (complex) optimum values. On the other hand, the
IPMS produces,
in general, a unique (real) optimum. 
In the following we will compare results
obtained with both methods.\\
We shall also consider for completeness another variation of the LDE-PMS
introduced in Ref. \cite{kleinert}, that we may call 
a ``renormalization group inspired" interpolation. This, more precisely,
modifies the simplest LDE substitution formula Eq. 
(\ref{subst}) by introducing consistently into the substitution 
the relevant critical exponent:
\begin{equation}
m \to m (1-\delta)^{1/\omega^\prime} \;\;,
\label{expcrit2}
\end{equation}
where
$\omega^\prime =
2\,\Omega/(2-\eta)$, $\eta$ being the anomalous dimension
of the critical propagator $\sim 1/p^{2-\eta}$
and $\Omega \equiv \beta^\prime(g_c)$, where 
$g_c$ is the critical coupling and $\beta(g)$ the renormalization group
beta function (see e.g. Ref. \cite{zustin}
for a review and definitions of the critical exponents). 
The benefit of the latter method in the present case is that the
value of the critical exponent is known exactly for large $N$:
$\omega^\prime (N\to\infty)=1$. (Incidentally,  
our more empirical IPMS approach introducing more parameters is  
in practice equivalent to the latter at second order in $\delta$, 
Eq. (\ref{subst2}), where the only additional parameter $a$ plays
thus a role similar to the critical exponent $\omega^\prime$). 
In our
numerical applications below, we shall thus consider
those three different forms of the LDE optimized perturbative expansion.

In order to resum the large-$N$ LDE type of series one should
now evaluate, in place of expression Eq. (\ref{basm0}), 
the equivalent contributions coming from 
the graphs shown in Figure 1 now including a mass term in
all propagators, including the plain tadpole term discussed in
the previous section
\begin{eqnarray}
r_c^{\rm tad}(m)&=& - \frac{Nu}{6}\int \frac{d^3\,p}{(2\pi)^3} \frac{1}{p^2+m^2}
\nonumber \\
&+&\frac { u N}{18}
\int \frac{d^3\,p}{(2\pi)^3} \;
\frac{d^3\,k}{(2\pi)^3} \;
\frac{1}{[{p}^2+m^2]^2}\;
\left [1 + \frac { u N}{6} B(k,m) \right ]^{-1}\;
\left[\frac{1}{({k} + {p})^2+m^2}-
\frac{1}{{k}^2+m^2 }\right ]   \;,
\label{basicN}
\end{eqnarray}
where the first term on the right hand side represents
the plain one-loop tadpole
while the series
represents the tadpole with bubble
insertions shown in Figure 1. However,
note carefully that
at the end the first term of the series   will not be considered since it does not
correspond to any 
real physical contribution being an artifact of the resummation as discussed, for
an equivalent
 series, in Ref. \cite {new}.
Now, the massive dressed propagator reads
\begin{equation}
B(k,m) = \int \;\frac{d^3
q}{(2\pi)^3}\;\frac{1}
{[q^2+m^2]\;[({k} + {q})^2+m^2]}
\;\; = \frac{1}{4\pi k}\;
 \arctan \left( \frac{k}{2m}\right )  \;,
\label{sigex}
\end{equation}
with $k \equiv |{\bf k}|$ and similarly for $p$, $q$ in Euclidean space. 
Contrary to the corresponding expression in the $m =0$ case
one cannot integrate exactly Eq. (\ref{basicN}) due to the
non-trivial dependence in $k$ and $m$ of
$B(k,m)$. However, the coefficients of the resulting perturbative
series at arbitrary orders can be calculated numerically by simple 
integrals \cite{new,braaten}. After discarding the spurious term discussed above
the result can be written as
\begin{equation}
r_c^{\rm tad}(m)= \frac {N\: u\: m}{24\pi} -
 {N\: u^2 }
\sum_{i=1}^\infty C_i \left ( - \frac { u N}{ m} \right )^i
\;, \label{exp20}
\end{equation}
where the perturbative coefficients are given by 
\begin{equation}
C_i=  \frac {1}{96 \pi^3} \left ( \frac {1}{48\pi} \right )^i
\int_0^{\infty} dz \frac {z^2}{(z^2+1)(z^2+4)}[A(z)]^i  \;,
\label{Js}
\end{equation}
with 

\begin{equation}
A(z) = \frac {2}{z} \arctan \frac {z}{2} \;,
\end{equation}
and $z=k/m$ (see e.g. Ref. \cite{new} for more details).
Similarly, the contributions of the exchange-type, shown in Figure 2, are
now to be evaluated with the massive propagator 
$B(k,m)$ in Eq. (\ref{sigex})
\begin{equation}
r_c^{\rm exc}(m)= -\Sigma(0)^{\rm exc }= - \int \frac{d^3\,k}{(2\pi)^3}
\frac{1}{[{k}^2+m^2]}\left ( \frac {u}{3} \right ) \left [ 1+ \frac {Nu
B(k,m)}{6} \right ]^{-1} \,\,.
\end{equation}
 Note that the first term of this series is nothing but
and additional plain tadpole contribution, however
with no overall $N$ factor. This is consistent, since we should in fact
consider for our purpose all contributions at the next-to-leading
$1/N$ order. Explicitly this terms gives a contribution $m\,u/(12\pi)$
to $r_c$, which combines with the first term of Eq. (\ref{exp20}) 
so that the final one-loop tadpole contribution 
to $r_c$ is $(N+2)\,m\, u/(24\pi)$,
which is consistent with the usual result
for arbitrary $N$ (see e.g. \cite{pra}) . 
Next, the second term corresponds to the only
divergent and scale
dependent contribution (``setting-sun") discussed above, 
which may be more conveniently evaluated
separately as shown 
in Refs. \cite {branieto,pra}.
All other higher order terms are finite and scale independent being easily
resummed.
Once more, the integral cannot be done
analytically but arbitrary order coefficients of
the  corresponding perturbative series are evaluated numerically 
and the final series can be cast into the form:
\begin{equation}
r_c^{\rm exc}(m)=
 \frac {N u^2}{18 (4\pi)^2}\left [\ln \left ( \frac{M}{m} \right )
+\frac{1}{2} +\ln(1/3) \right ]
-m\:u\; \sum_{i=2}^\infty D_i \left ( - \frac { u N}{ m} \right )^i
\;, \label{exp21}
\end{equation}
where  
\begin{equation}
D_i=  \frac {1}{6 \pi^2} \left ( \frac {1}{48\pi} \right )^i
\int_0^{\infty} dz \frac {z^2}{(z^2+1)}[A(z)]^i  \;.
\label{DJs}
\end{equation}

As expected,
the scale dependence ($\ln(M)$) within the perturbative
result Eq. (\ref{exp21}) above again has the very same coefficient
than the corresponding scale dependence of the non-perturbative
result, Eq. (\ref{rcexc}).
Note also that our selection of graphs takes automatically care of the HP theorem
with $r_c$ being 
appropriately replaced. It is easy to see that the order-$\delta^2$ result
obtained 
in the previous sub-section can be immediately reproduced by our resummed type of
calculation.
The final total $O(N^{-1})$ contribution is then obtained by adding 
Eqs. (\ref {exp20})
and (\ref {exp21}) (with however the consistent 
replacement $N \to N+2$
in the first one-loop tadpole term of Eq.   (\ref {exp20}), as discussed
above).

\subsection{ The large-$N$ case and convergence of the LDE}

Let us now perform the OPMS-LDE and the IPMS-LDE substitutions discussed above  in
order
to obtain optimized results also discussing the convergence properties of each
method.
Actually, as discussed
above in section IV.B, the first series given by Eq. (\ref{exp20})
is equivalent,
up to a trivial overall factor $-u/6$, to the series defining
the quantity $k$ relevant to the $\Delta T_c/T_c$ shift in the 
critical temperature. For the latter series, 
a good numerical convergence
towards the exact result has already been obtained in the case
of the IPMS \cite{knp}. On the other hand, 
the exact analytical result for the 
chain-exchange type of graphs, Fig. 2, is unambiguously defined
separately as Eq. (\ref{rcexc}).    
We can thus for numerical analysis
simplicity discard in a first stage the part of the perturbative 
contributions Eq. (\ref{exp20}) to concentrate on the
other non-trivial series Eq. (\ref{exp21}) in order
to examine its eventual convergence properties in the framework of the
OPMS and/or IPMS. Since we shall also consider in a second stage 
the complete series, it provides in this way two numerically 
independent analysis of the (eventual) convergence properties of the
different PMS methods. 
Accordingly the exact result to
which the resummation of Eq. (\ref{exp21}) alone should be compared 
with, is the second term in Eq. (\ref{rcNex}), which numerically is
equal to $\sim 0.00181497 \, u^2$.\\
We give the results of IPMS and standard OPMS
in Table I, second and fourth column respectively. The
third column indicates the corresponding value of the 
IPMS leading parameter\footnote{Similarly to the results in Ref.
\cite{knp}, for the real IPMS optimum solutions higher order
interpolation parameters 
turn out to be numerically much smaller in magnitude than $a$, so
that we omit to give their precise values here.}  $a$, 
cf. Eq. (\ref{subst2}).   
Next, we give in the last right column the
results obtained when considering, following Refs. \cite{kleinert,
kastening2}, the renormalization group inspired
interpolation method (that
we shall dub RGPMS for comparison), as defined in Eq. (\ref{expcrit2}).
Notice that at second order ($k=2$), the IPMS and OPMS result coincide
simply because it is not possible to implement an (independent) new 
interpolation parameter, so that the IPMS actually starts at order three.
(The empty RGPMS result entry at second order is simply due to the
fact that for $\omega^\prime =1$ there is an accidental exact cancellation
of the linear in $m$ (tadpole)  term of Eq. (\ref{exp20}), such that
the RGPMS non-trivial optimization solutions also start at order three).

One can see that at first rough sight,  
 all different PMS approaches seem to perform very well, all
exhibiting a numerically convergence behaviour towards the exact result   
as the order increases. Moreover, 
all the results are already remarkably good at third order.  
In practice the IPMS method, introducing one
more parameter at each successive order, becomes quite cumbersome 
at large orders, so
that we somewhat arbitrarily stopped our analysis at order twelve, 
beyond which it became unreasonably time consuming for our calculation
(performed with Mathematica \cite{mathematica}).
When examining in more details the behavior of each different method,
we first observe that the (real part of) the
standard OPMS method results get 
very close to the exact
result at larger and larger orders,  
letting aside the embarassing non-zero but small
imaginary parts. However, in Table I we have
selected among several (many at large orders) real and complex optimal
results those closest to the known exact result. Nevertheless, if 
selecting instead only the real solutions 
we find a slightly less convergent series of approximations, but still
quite satisfactory (the real solutions are provided in Table I, 
whenever they exist).
Alternatively, the
IPMS has the advantage that it 
gives at any order a unique real optimal solution, and
seemingly converging somewhat more slowly but regularly
towards the expected result.       
As far as the RGPMS method is concerned,
some of the optimization results appear at large orders to be
even better that the two previous ones, being  
impressively close to the exact one
as the perturbative order increases. This may be due to the fact
that the known large-$N$ exact value $\omega^\prime=1$ enters this
interpolation. However, the RGPMS suffers a priori from the same
drawback than the ordinary PMS, namely that the 
real optimal solutions are not unique, 
as explicitly illustrated in the table, and we
have also selected the best among many more (mainly
complex) optimized  
solutions. It is interesting to remark finally that, as observed
similarly in ref. \cite{knp}, as the order increases
 the leading IPMS parameter $a$ seems to approach (though very slowly) 
a critical value $a=2$ (which, upon neglecting other higher order
IPMS interpolation parameters, would
correspond to $\omega^\prime=1$ in the RGPMS approach). \\   
\begin{table}[htb]
\begin{center}
\caption{IPMS, OPMS and RGPMS results for the large-$N$ BEC
 $r_c \equiv r^{\rm exc}_c $
at different orders $k$. $r_c^{(N,{\rm exc})}=
0.00181497\ldots u^2$} 
\begin{tabular}{|c||c|c||c||c|}
\hline
$ k $ &  $r_c^{IPMS}/u^2$ & $ a $& $r_c^{best\: PMS}/u^2$ 
& $r_c^{best\:RGPMS}/u^2\:(\omega^\prime =1)$ 
 \\ \hline\hline
2   &  0.00206 &       & 0.00206 &   \\
\hline
3  & 0.001845 & 1.43     & $0.001845$  &    0.001829            \\
\hline
4 & 0.001834 &  1.43 &   $0.001809\pm 3.4 \:10^{-5}I,
\;0.001857 $  & 0.0018199  \\
\hline
5 &  0.001830  & 1.49 &  $0.001809\pm 1.6 \:10^{-5}I$    &  0.001821 \\
\hline
6 &  0.001827  & 1.53 &  $0.001813\pm 9. \:10^{-6}I,
\; 0.001837$    & 0.0018195 \\
\hline
7 &  0.001825  & 1.57 &   $0.001817\pm 6.6 \:10^{-6}I$   & 
$0.001816\pm 2. \:10^{-7}I,\; 0.0018189$ \\
\hline
8 & 0.001824   & 1.60 &  $0.001820\pm 6.7 \:10^{-6}I,\;
0.001830 $  &  0.0018155, 0.0018185 \\
\hline
9 & 0.001823  &   1.63 &   $ 0.001820\pm 1.0 \:10^{-5}I$ & 
$0.0018154\pm 9. \:10^{-8}I,\; 0.0018182$\\
\hline
10 & 0.001822  &   1.65 &  $0.001819\pm 1.\: 10^{-5}I,\;
0.001827$   & 0.0018153, 0.0018180 \\
\hline
11 & 0.0018213  &   1.67 &  $0.0018199\pm 9.\: 10^{-6}I$   & 
$0.0018152\pm 5.\: 10^{-8}I,\; 0.0018179$ \\
\hline
12 & 0.0018208  &   1.69 &  $0.0018199\pm 9.\: 10^{-6}I,\;
0.0018252 $   &0.00181513, 0.00181776\\
\hline
20 &  &  &     $0.0018178 \pm 1.\: 10^{-5}I,\;0.0018219 $ & 
0.00181509, 0.0018173 \\
\hline
\end{tabular}
\end{center}
\end{table}
Next, as a second stage independent numerical analysis using the same
OPMS, IPMS or RGPMS methods, we consider  the full perturbative
series ($r^{\rm tot}_c $)
 consisting of the sum of expressions defined in
Eq.
(\ref{exp20}) and (\ref{exp21}), respectively. The results of the
optimization are shown in Table II. The numerical behaviour 
appears to be similar to the one in Table I, namely with 
all different PMS methods appearing to converge quite well.
However, we stress once more the uniqueness of the real
IPMS optimized solutions at each orders, exhibiting thus a relatively  
slower but completelly unambiguous apparent convergence for this method.  
Finally, if neglecting the imaginary parts of the
optimized solutions (which are however very small in this case),  
the RGPMS approach appears again to give the best results at large
orders, though real solutions do not always exist at arbitrary orders,
just like in the OPMS case.     
\begin{table}[htb]
\begin{center}
\caption{IPMS, OPMS and RGPMS
results for the large-$N$ BEC
 $r^{\rm tot}_c $
at different orders $k$. $r_c^{(N)}=
0.00216675\ldots u^2$}
\begin{tabular}{|c||c|c||c||c|}
\hline
$ k $ &  $r_c^{IPMS}/u^2$ & $ a $& $r_c^{best\: PMS}/u^2$ 
& $r_c^{best\:RGPMS}/u^2\:(\omega^\prime=1)$
 \\ \hline\hline
2   &  0.002063 &       & 0.002063 &   \\
\hline
3  & 0.001973 & 1.14     &  $0.001960  \pm 7.9\: 10^{-6} I$& 0.002060 \\
\hline
4 & 0.002041 &  1.70 & 0.001991 &
$0.002101 \pm 2.1\: 10^{-5} I$ \\
\hline
5 &  0.002065  & 1.73 &  $0.002020, 0.001995 $    &   0.0021254   \\
\hline
6 &  0.002081  & 1.75 &  $0.002033 \pm 3.4 \: 10^{-5}I,\;
 0.001999$   &
$0.002135 \pm 7.5\: 10^{-6}I$ \\
\hline
7 &  0.002093  &  1.76 &   $0.002046,0.002001 $   & 0.0021422 \\
\hline
8 & 0.002101   & 1.78 &  $0.002055 \pm 1.6 \: 10^{-4}I,\;
0.002002$  &
$0.0021463 \pm 3.7 \:10^{-6}I$  \\
\hline
9 & 0.002108  &   1.79 &   $0.002067 \pm 1.7  \:10^{-4}I,\;
0.002061$ & 0.0021495 \\
\hline
10 & 0.002114  &   1.80 &  $0.002076 \pm 1.7 \: 10^{-4}I,\;
0.002061$ &
$0.0021517 \pm 2.2\: 10^{-6}I$  \\
\hline
11 & 0.002118  &   1.81 &  $0.002083 \pm  1.7 \: 10^{-4}I,\;
0.002071$  & 0.0021535\\
\hline
12 & 0.002122  &   1.82 &   $0.002089 \pm  1.8\:  10^{-4}I,\;
0.002006$ &
$0.0021549 \pm 1.5\: 10^{-6}I$ \\
\hline
20 &   &    &    $0.002121 \pm  1.3\:  10^{-4}I,\;0.002008$ &  
$0.0021605 \pm 7\:10^{-7}I$ \\
\hline
\end{tabular}
\end{center}
\end{table}
\subsection{ The finite $N=2$ case}

Here, for completeness, we extend some of 
the results of Ref. \cite {knp} for $r_c$ at 
arbitrary $N$ by incorporating the 
very recently available improved six-loop and seven-loop relevant
perturbative contributions performed in Refs.
\cite{kastening2, newkast, private}. One obtains 
to order-$u^7$ and for  $N=2$

\begin{equation}
r_c^{(5)}= - \Sigma_{\rm ren}^{(5)}(0)=  \frac {u m}{6\pi}
+ u^2 A_2 \left [ \ln
\left ( \frac{M}{m} \right ) +\frac{1}{2} +\ln \frac{1}{3} \right ]
+ m^2 \sum_{i=3}^7 \left ( \frac {- u }{m} \right )^i A_i \;\;,
\label{rcNF}
\end{equation}
where in our normalization,
 $A_2=1.40724 \times 10^{-3}$, $A_3=8.50891 \times
10^{-5}$, $A_4=3.46941 \times 10^{-6}$,
$A_5= 2.23296 \times 10^{-7}$, 
$A_6= 1.81417 \times 10^{-8}$, and
$A_7= 1.71627 \times 10^{-9}$. 
Performing  the IPMS-LDE and OPMS-LDE substitutions, 
already discussed, one 
obtains the optimized LDE results   shown in Table III 
below \footnote{The slightly different results at orders 3-5 
of the IPMS method for 
$c^{\prime\prime}_2$ with respect to Table VIII of Ref.\cite{knp}, are
simply due to our updated use of the most recent analytical and  
numerical improvements 
in the determination of the perturbative series coefficients 
given in Ref.
\cite{newkast}. Note that $c^{\prime\prime}_2$ cannot be determined
at order $\delta^7$ since this requires the use of some eight-loop contributions
to $\langle \phi^2 \rangle$ which are not available at present.}
for $M=u/3$.
 As one can see, the results of our alternative IPMS method
show a remarkable
agreement with the numerical lattice result. Concerning the results
of the standard OPMS method, as shown in the right column of Table III,
 they approach closely the lattice
result at third order  while the fourth order   result departs from the
LS value. Then, the
fifth and higher orders prediction approache the LS result again,  
considering again here the real parts of complex optimal solutions  
(although the imaginary parts are very small in this case).\\ 
\begin{table}[htb]
\begin{center}
\caption{IPMS
versus standard OPMS results for the physical $N=2$ BEC
 $r_c $, and correspondingly $c^{\prime\prime}_2$ 
at different orders $k$. $r_c^{\rm LS}=
(0.0019201 \pm 0.0000021)\, u^2$}
\begin{tabular}{|c||c|c|c||c|}
\hline
$ k $ &  $r_c^{IPMS}/u^2$& $c^{\prime\prime}_2(IPMS)$ & 
$ a $& $r_c^{best \;PMS}/u^2$
 \\ \hline\hline
2   &  0.00315 &  101.2 &    & 0.00315    \\
\hline
3  & 0.001828 & 69.85 & 1.77   &  $0.00215  \pm 7.8\: 10^{-4} I$\\
\hline
4 & 0.0018988 &  78.31 & 2.02  & $0.00247$  \\
\hline
5 &  0.0018992 & 74.44 & 2.116 &   $0.00226  \pm 3.3\: 10^{-4} I$     \\
\hline
6 &  0.0019034 & 75.51 & 2.183 &   $0.00205  \pm 4.5\: 10^{-4} I$     \\
\hline
7 &  0.0019026  & & 2.237 &   $0.001887  \pm 2.0\: 10^{-4} I$     \\
\hline
\hline
\end{tabular}
\end{center}
\end{table}

\section{Conclusions}

Considering an effective three-dimensional 
model for homogeneous dilute weakly
interacting Bose 
gases we have evaluated
the physical quantity $r_c$ which is directly related to the critical chemical
potential. Our evaluations made use
of the standard $1/N$ approximation as well as 
different variations of the optimized perturbation (linear $\delta$
expansion) method. 
As discussed in the text, this quantity is important in the determination of a
numerical coefficient
($c_2^{\prime \prime}$)
appearing at order-$a^2 n^{2/3}$ in the expansion of $\Delta T_c$. Surprisingly,
the large-$N$ result
$r_c \simeq 0.00216672 \, u^2$ compares well with the most recent lattice
simulation result, 
$r_c^{\rm LS} = (0.0019201 \pm 0.0000021)\,u^2$ \cite {arnoldrc,
second} even for the case $N=2$. As a matter of
fact the large-$N$ approximation seems to work better for the determination of
$r_c$ than for the 
 $\langle \phi^2 \rangle$ prediction performed in Ref. \cite {baymN}. Then, having
established the large-$N$ result
we have used the linear $\delta$ expansion, in the large-$N$ limit, to
examine its convergence properties, in a way similar as it
was done in a previous work  \cite {new} for the quantity $\langle \phi^2
\rangle$.
Our analysis shows that the standard
optimization procedure produces converging results which can be better controlled
if one uses a recently improved version of
the method \cite {knp}. 
In this way one is able to obtain, at each order in
$\delta$, only one real
optimized solution,
contrary to what happens if one uses the standard optimization procedure.
We also obtained that another variation \cite{kleinert} of the
LDE, introducing consistently the critical exponent $\omega^\prime=1$
performs very well in the large $N$ limit.  
Apart from determining the large-$N$ result our work shows  the reliability of the
LDE finite $N$ results for $r_c$ obtained in Ref. \cite {knp}.
Finally, considering the  large-$N$ \cite {baymN} and IPMS-LDE \cite {knp} 
results for $\kappa$, one can  establish the analytical values for
$c_2^{\prime \prime}$. For $N=2$ they are
$c_2^{\prime \prime}\simeq 95.79$ (large-$N$ approximation). Also, our 
numerical results in Table III for 
 $c_2^{\prime \prime}$ from the IPMS-LDE at the latest available 
orders ($\delta^5, \delta^6$) compare very well 
with the lattice simulation result
$c_2^{\prime \prime} = 75.7 \pm 0.4$ \cite {arnoldrc, second}.\\
{\em Note added:} After completing a first version of
this paper (eprint archive: cond-mat/0408538, v1) we became aware of 
the very recent work by Kastening \cite{newkast}, where an 
analytical evaluation of $r_c$ in the large $N$ limit is also performed
as well as the relevant perturbative contributions up to seven loop for  
arbitrary $N$  (partly using some previously unpublished
seven-loop material by B. Nickel), 
with a corresponding new evaluation of $r_c$  
in the framework of the ``variational perturbation theory" (VPT). 
Our large $N$ exact result for $r_c$, Eq. (\ref{rcNex}), 
maybe be considered as an independent
cross-check of his result, since ours is performed with a different
method. Concerning his numerical VPT results
for $N=2$, at seven-loop order 
they lie in the range\footnote{Note for comparison a trivial
factor of 4 (for $N=2$) difference in the 
normalization of $r_c/u^2$
in ref. \cite{newkast} with
respect to us or LS results}: 
$r_c/u^2 \sim 0.0019028-0.0019401$, thus  
appearing very consistent with our results in Table III.  

\acknowledgments

M.B.P. is partially supported by CNPq-Brazil.
We thank Boris Kastening for useful communication relevant to
the six loop evaluation of $r_c$ for $N=2$. M.B.P. also thanks  the
LPMT (Universit\'{e} de Montpellier II) for a CNRS guest grant.

\end{document}